\documentclass{article}
\usepackage{amsfonts}
\usepackage{graphicx}
\usepackage{amsmath}

\newtheorem{theorem}{Theorem}

\newtheorem{proposition}[theorem]{Proposition}

\input{tcilatex}

\begin{document}

\title{\textbf{CANCELLATION OF VORTICITY IN STEADY-STATE NON-ISENTROPIC
FLOWS OF COMPLEX FLUIDS}}
\author{\textbf{Paolo Maria Mariano} \\
Dipartimento di Ingegneria Strutturale e Geotecnica\\
Universit\`{a} di Roma ``La Sapienza''\\
via Eudossiana 18\\
00184 Roma (Italy)\\
e-mail: paolo.mariano@uniroma1.it}
\maketitle

\begin{abstract}
In steady-state non-isentropic flows of perfect fluids there is always
thermodynamic generation of vorticity when the difference between the
product of the temperature with the gradient of the entropy and the gradient
of total enthalpy is different from zero. We note that this property does
not hold in general for complex fluids for which the prominent influence of
the material substructure on the gross motion may cancel the thermodynamic
vorticity. We indicate the explicit condition for this cancellation
(topological transition from vortex sheet to shear flow) for general complex
fluids described by coarse-grained order parameters and extended forms of
Ginzburg-Landau energies. As a prominent sample case we treat first
Korteweg's fluid, used commonly as a model of capillary motion or phase
transitions characterized by diffused interfaces. Then we discuss general
complex fluids. We show also that, when the entropy and the total enthalpy
are constant throughout the flow, vorticity may be generated by the
inhomogeneous character of the distribution of material substructures, and
indicate the explicit condition for such a generation. We discuss also some
aspects of unsteady motion and show that in two-dimensional flows of
incompressible perfect complex fluids the vorticity is in general not
conserved, due to a mechanism of transfer of energy between different levels.
\end{abstract}

\section{Introduction and statement of the problem.}

For steady-state non-isentropic flows of perfect fluids, the following
pointwise relation holds throughout the regular region $\mathcal{B}$ of the
three-dimensional Euclidean point space $\mathcal{E}$ occupied by the fluid: 
\begin{equation}
\mathbf{\omega }\times \mathbf{v}=\vartheta grad\eta -gradH,  \label{1}
\end{equation}%
where $\mathbf{v}$ is the velocity of each material element $\left( \mathbf{%
v=\tilde{v}}\left( \mathbf{x}\right) \right) $, $\mathbf{\omega }$\ the
vorticity $\left( \mathbf{\omega =}\emph{curl}\mathbf{v}\right) $, $%
\vartheta $\ the temperature, $\eta $\ the entropy density, $H$\ the total
enthalpy density ($grad$ and $curl$ are calculated with respect to $\mathbf{%
x\in }\mathcal{B}$). Equation (1) is known as Crocco-Vaszony's theorem (see
[1-3]). It states basically that we may observe thermodynamical generation
of vorticity when 
\begin{equation}
\vartheta grad\eta -gradH\neq 0.  \label{2}
\end{equation}

For complex fluids, for which the gross behavior is influenced by a
prominent presence of material substructure, the condition (2) is no more
sufficient to assure vorticity and we may have situations in which the
effects due to the substructure cancel the thermodynamic generation of
vorticity, or may produce it even when the entropy and the total enthalpy
are constant.

In what follows we give explicit conditions for the occurrence of such a
topological transition, due to the complex structure of the fluid, from a
vortex sheet to a shear flow and vice versa\footnote{%
Such a kind of transition may occur in disparate circumstances, as, e.g.,
geophysical flows [18].}. To fix basic ideas, we consider first the simplest
case of complex fluid, the one of Korteweg, in which effects associated with
the gradient of the specific volume are accounted for. Then, we treat a
general case (following the unified theory in [4]) in which each material
element at $\mathbf{x}$, at the time $t$, has an articulated morphology and
can be considered as a `system', described by a coarse grained order
parameter $\mathbf{\nu }$, taken here as an element of an `abstract'
differentiable manifold $\mathcal{M}$ to cover a class of complex fluids as
large as possible. The possible non-uniform distribution of the order
parameters influence the formation of vorticity during the flow. Finally, we
discuss the unsteady case and show that in a two-dimensional flow of
incompressible perfect complex fluids the vorticity is in general \emph{not}
conserved (as in simple perfect fluids) unless an explicit condition is
satisfied. The influence of the material substructure, rendering not
conserved the vorticity, is not of viscous nature: there is just a transfer
of energy from the coarse grained (macroscopic) level to the substructural
one, altering the macroscopic energetic landscape of the vorticity (that may
generate possible drag reduction in turbulent flows). This phenomenon
implies the possibility that even in a two-dimensional flow of complex
fluids a vortex may be destroyed and, possibly, reconnected in two or more
subvortices.

For the sake of simplicity we do not consider applied external fields.

\section{The case of Korteweg's fluid.}

Motivated by the work of van der Walls and previous ideas of Lord Rayleigh
and Gibbs, to model at coarse scale the effects of interfacial tension in
capillary phenomena, in 1901, Korteweg [5] introduced a model of fluid in
which the energy depends on the gradient of the density. Experimental works
have supported this approach [6-8] and the model has been source of
analytical investigations [9, 10]. A detailed critical review about this
topic can be found in [11]. The work of Korteweg has been the germ of a
general modeling of `materials of grade $N$' in which higher order
deformation gradients up to grade $N$ are involved in constitutive equations
(see [12] for the general thermodynamic treatment). In 1985 Capriz [13]
proved clearly that the constitutive dependence on higher order
`deformation' gradients is an indicator of the prominent presence of a
complex material substructure which needs the introduction of some order
parameter, under the assumption of the existence of an internal constraint
linking the order parameter with the deformation (Korteweg's model follows
as a special case). Our prominent interest in the present setting is not
properly on capillary phenomena, the original motivation of Korteweg's
model, rather it is focused on the circumstance that such a model is
currently used to represent phase transitions on the basis of a \emph{%
diffuse interface} approach (see [14] for a detailed review). In its
simplest form, in fact, the free energy is of Cahn-Hilliard type: the
gradient of the mass density account for the free energy excess of the
interfacial region between phases. Such an excess of energy may influence
the generation of vorticity in non-isentropic flows: it may cancel or
produce vorticity.

\subsection{The main result about Korteweg's fluids.}

Here we analyze the simplest case of complex fluid, i.e., the perfect
Korteweg's one: a compressible perfect fluid characterized by the functional
dependence of the potential $\phi $ on the gradient $grad\iota $ of the
specific volume $\iota $ (being $\iota =\tilde{\iota}\left( \mathbf{x}%
\right) =\det \mathbf{g}\left( \mathbf{x}\right) $, with $\mathbf{g}\left( 
\mathbf{x}\right) $ the metric in space at $\mathbf{x}\in \mathcal{B},$ and $%
grad$ is calculated with respect to $\mathbf{x}$) in addition to $\iota $
itself; in other words, we consider the case in which $\phi =\phi \left(
\iota ,grad\iota ,\eta \right) $, with $\phi $ a sufficiently smooth
function.

In the following, we write formally $\phi \left( \iota ,grad\iota ,\eta
\right) $ being conscious that since $\phi $ is a scalar function depending
on a vector (namely $grad\iota $), for reasons of objectivity (frame
indifference under the action of $SO\left( 3\right) $) $\phi $ must depend
on the vector $grad\iota $ only through its amplitude $\left\vert grad\iota
\right\vert $. The simplest choice of $\phi $ is Cahn-Hilliard's form: $\phi
=f\left( \iota ,\eta \right) +\frac{1}{2}\beta \left\vert grad\iota
\right\vert ^{2}$, with $\beta $\ some constant and $f$ a function
describing, e.g., a two-well energy; then $\phi $ may account for the
spinodal decomposition. We also indicate with $\rho $ the current density of
mass of the fluid and consider the referential density $\rho _{0}$ equal to $%
1$ so that $\iota =\rho ^{-1}$. In general, the Lagrangian density of the
Korteweg's fluid is considered to be given by%
\begin{equation}
\frac{1}{2}\rho \left\vert \mathbf{v}\right\vert ^{2}+\chi \left( \iota ,%
\dot{\iota}\right) -\rho \phi \left( \iota ,grad\iota ,\eta \right)
\label{3}
\end{equation}%
for non-isentropic flows, where $\chi $\ is a sufficiently smooth function
such that $\partial _{\dot{\iota}\dot{\iota}}^{2}\chi \neq 0$, $\chi \left(
\iota ,\dot{\iota}\right) \geq 0$ and the equality sign holds when $\dot{%
\iota}=0$. The function $\chi $ accounts for possible substructural inertia
effects that may generate gradient inertia. In other words, the appearance
of the time derivative of $\iota $, namely $\dot{\iota}$, in (\ref{3}) is 
\emph{not} of viscous nature, rather it is of inertial nature and the
physical dimensions of $\chi $ are the ones of a kinetic energy. Notice that
the kinetic energy is the partial Legendre transform with respect to $\dot{%
\iota}$\ of $\chi $\ which is called kinetic \emph{co-energy}, following the
terminology of [4]. The possible existence of gradient inertia effects in
the Korteweg's fluid has been pointed out in [13], however, these effects
seem to be negligible unless the fluid oscillates at very high frequencies.
In any case we formally account for the presence of $\chi $ in the
following, because in the general setting treated below its effects may be
sensible.

Hereafter, the dot indicates the scalar product associated with the natural
dual pairing; moreover, we denote with $\partial _{y}$ the partial
derivative with respect to the argument $y$ and, for different types of
linear operators, with a superscript $T$ \ we indicate the transposition
rendered explicit by a subsequent detailed expression in components, when
the meaning of the transposition is not immediately evident. For $A$ and $B$
tensors of the same order, $A\cdot B$ indicates scalar product.

The main result of the present section is summarized in the proposition and
the corollaries below.

\begin{proposition}
For a steady-state non-isentropic flow of a Korteweg's fluid, the following
pointwise relation holds:%
\begin{eqnarray}
\mathbf{\omega }\times \mathbf{v} &=&\vartheta grad\eta -gradh-grad\left(
\iota ^{2}div\rho \partial _{grad\iota }\phi +\partial _{grad\iota }\phi
\cdot grad\iota \right) +  \notag \\
&&+\iota grad\left( \frac{d}{dt}\left( \partial _{\dot{\iota}}\chi \right)
-\partial _{\iota }\chi \right) .  \label{4}
\end{eqnarray}
\end{proposition}

In equation (\ref{4}), $h$ is the total enthalpy density of the Korteweg's
fluid and its explicit expression is given in Section 2.3.

\textbf{Corollary 1}. \emph{In a perfect Korteweg's fluid, the gradient
effects cancel the thermodynamic generation of vorticity when}%
\begin{equation*}
\vartheta grad\eta -gradh=grad\left( \iota ^{2}div\rho \partial _{grad\iota
}\phi +\partial _{grad\iota }\phi \cdot grad\iota \right) -
\end{equation*}%
\begin{equation}
-\iota grad\left( \frac{d}{dt}\left( \partial _{\dot{\iota}}\chi \right)
-\partial _{\iota }\chi \right) .  \label{5}
\end{equation}

\textbf{Corollary 2}. \emph{In the case in which both the specific entropy
and the total enthalpy are constant throughout the flow of a perfect
Korteweg's fluid, the gradient effects may generate vorticity:}%
\begin{equation}
\mathbf{\omega }\times \mathbf{v}=-grad\left( \iota ^{2}div\rho \partial
_{grad\iota }\phi +\partial _{grad\iota }\phi \cdot grad\iota \right) +\iota
grad\left( \frac{d}{dt}\left( \partial _{\dot{\iota}}\chi \right) -\partial
_{\iota }\chi \right) .  \label{6}
\end{equation}

\subsection{Preliminaries to perfect Korteweg's fluids.}

\subsubsection{Equilibrium.}

In absence of body forces, the equilibrium of the fluid in $\mathcal{B}$ can
be determined by evaluating the variation of the functional%
\begin{equation}
\Phi =\int_{\mathcal{B}}\rho \phi \left( \iota ,grad\iota \right)  \label{7}
\end{equation}%
and putting it equal to zero, i.e.%
\begin{equation}
\hat{\delta}\int_{\mathcal{B}}\rho \phi \left( \iota ,grad\iota \right) =0.
\label{8}
\end{equation}%
for any choice of $\delta \mathbf{x}$, where $\hat{\delta}$ indicates the 
\emph{total} variation (in the sense used in elementary treatises when one
deals with `total' time derivatives), so that we have (see [4], p. 31)%
\begin{eqnarray}
\hat{\delta}\int_{\mathcal{B}}\rho \phi &=&\int_{\mathcal{B}}\delta \left(
\rho \phi \right) +\int_{\partial \mathcal{B}}\rho \phi \mathbf{n\cdot }%
\delta \mathbf{x=}  \notag \\
&=&\int_{\mathcal{B}}\left( \delta \left( \rho \phi \right) +div\left( \rho
\phi \delta \mathbf{x}\right) \right) =\int_{\mathcal{B}}\rho \hat{\delta}%
\phi ,  \label{9}
\end{eqnarray}%
where $\mathbf{n}$ is the outward unit normal to the boundary $\partial 
\mathcal{B}$ of $\mathcal{B}$.

We note that%
\begin{equation}
\hat{\delta}\phi \left( \iota ,grad\iota \right) =\partial _{\iota }\phi 
\hat{\delta}\iota +\partial _{grad\iota }\phi \hat{\delta}grad\iota ,
\label{10}
\end{equation}%
where $\hat{\delta}\iota $\ and $\hat{\delta}grad\iota $\ are the total
variations of the relevant arguments calculated taking into account that $%
\iota $ varies in space. We have in fact%
\begin{equation}
\hat{\delta}\iota =\iota div\delta \mathbf{x,}  \label{11}
\end{equation}%
as a consequence of Euler formula, and by chain rule%
\begin{equation}
\hat{\delta}grad\iota =grad\hat{\delta}\iota -\left( grad\delta \mathbf{x}%
\right) ^{T}grad\iota  \label{12}
\end{equation}%
(see [4] for other details). By developing the variation in (\ref{8}),
taking into account (\ref{10})-(\ref{11}) and using repeatedly the Gauss
theorem, we recognize that the arbitrariness of $\delta \mathbf{x}$ implies
in the bulk the pointwise balance%
\begin{equation}
div\mathbf{T}=\mathbf{0,}  \label{13}
\end{equation}%
(as Euler-Lagrange equation) with%
\begin{equation}
\mathbf{T}=-p\mathbf{I-T}^{E},  \label{14}
\end{equation}%
where $\mathbf{I}$ is the second-order unit tensor, $p$ a non-standard
pressure given by%
\begin{equation}
p=-\rho \iota \partial _{\iota }\phi +\iota div\rho \partial _{grad\iota
}\phi ,  \label{15}
\end{equation}%
and $\mathbf{T}^{E}$ a \emph{non-viscous} stress of Ericksen's type called
in this context Korteweg stress, i.e.%
\begin{equation}
\mathbf{T}^{E}=grad\iota \otimes \rho \partial _{grad\iota }\phi .
\label{16}
\end{equation}

\subsubsection{Flows.}

Let us consider first isentropic flows. The dynamics of a Korteweg's fluid
can be simply derived by making use of a Hamilton variational principle of
the form%
\begin{equation}
\hat{\delta}\int_{0}^{\bar{t}}\int_{\mathfrak{b}}\left( \frac{1}{2}%
\left\vert \mathbf{v}\right\vert ^{2}+\chi \left( \iota ,\dot{\iota}\right)
\right) -\int_{0}^{\bar{t}}\text{ }\hat{\delta}\int_{\mathfrak{b}}\rho \phi
\left( \iota ,grad\iota \right) =0,  \label{17}
\end{equation}%
where the time $t$ runs in $[0,\bar{t}]$ and $\hat{\delta}$\ means \emph{%
total} variation as before. By developing (\ref{17}), we realize that the
arbitrariness of $\delta \mathbf{x}$ implies the standard balance%
\begin{equation}
\rho \frac{d\mathbf{v}}{dt}=div\mathbf{T.}  \label{18}
\end{equation}%
The stress tensor $\mathbf{T}$ is of the form%
\begin{equation}
\mathbf{T}=-\bar{p}\mathbf{I-T}^{E},  \label{19}
\end{equation}%
but now the non-standard pressure $\bar{p}$\ accounts for kinetic pressure
effects and is%
\begin{equation}
\bar{p}=-\rho \iota \partial _{\iota }\phi +\iota div\rho \partial
_{grad\iota }\phi -\rho \iota \frac{d}{dt}\partial _{\dot{\iota}}\chi
+\partial _{\iota }\chi ,  \label{20}
\end{equation}%
while $\mathbf{T}^{E}$\ remains the same of (\ref{16}).

\subsection{Proof of Proposition 1.}

We denote with $q$ the amplitude of the velocity, i.e., $q=\left\vert 
\mathbf{v}\right\vert $. Then, by using the standard Lagrange formula%
\begin{equation}
\frac{d\mathbf{v}}{dt}=\partial _{t}\mathbf{v+}\left( grad\mathbf{v}\right) 
\mathbf{v}=\partial _{t}\mathbf{v+\omega }\times \mathbf{v+}\frac{1}{2}%
gradq^{2},  \label{21}
\end{equation}%
we recognize that in the case of \emph{steady motion}, i.e., when $\partial
_{t}\mathbf{v=0}$, the balance equation (\ref{18}) can be written as%
\begin{equation}
\mathbf{\omega }\times \mathbf{v}=-\frac{1}{2}gradq^{2}-\iota grad\bar{p}%
-\iota div\mathbf{T}^{E}.  \label{22}
\end{equation}

When the flow is not isentropic, we may identify the potential $\phi $ with
the internal energy and write%
\begin{equation}
\phi =\phi \left( \iota ,grad\iota ,\eta \right) ,  \label{23}
\end{equation}%
being $\eta $\ the entropy density. From (\ref{23}), we define the \emph{%
specific enthalpy density} $\xi $ as the opposite of the partial Legendre
transform of $\phi $\ with respect to the kinematic variables, so that we
have%
\begin{equation}
\xi =\phi -\iota \partial _{\iota }\phi -\partial _{grad\iota }\phi \cdot
grad\iota ,  \label{24}
\end{equation}%
i.e., from (\ref{20}),%
\begin{equation}
\xi =\phi +\iota \bar{p}-\iota ^{2}div\rho \partial _{grad\iota }\phi +\iota 
\check{p}-\partial _{grad\iota }\phi \cdot grad\iota ,  \label{25}
\end{equation}%
where%
\begin{equation}
\check{p}=\rho \iota \frac{d}{dt}\left( \partial _{\dot{\iota}}\chi \right)
-\partial _{\iota }\chi  \label{26}
\end{equation}%
is the kinetic pressure due to possible gradient inertia effects. By
indicating with $\vartheta $ the temperature given by $\partial _{\eta }\phi 
$, the calculation of the gradient of (\ref{25}) allows us to obtain%
\begin{eqnarray}
-\iota grad\bar{p} &=&\vartheta grad\eta -grad\xi +\iota grad\check{p}-\iota
^{2}grad\left( div\rho \partial _{grad\iota }\phi \right) -  \notag \\
&&-\left( grad\partial _{grad\iota }\phi \right) ^{T}grad\iota -\iota \left(
div\rho \partial _{grad\iota }\phi \right) grad\iota .  \label{27}
\end{eqnarray}%
By inserting (\ref{27}) in (\ref{22}), taking into account that%
\begin{equation}
div\mathbf{T}^{E}=\left( div\rho \partial _{grad\iota }\phi \right)
grad\iota +\left( \rho \partial _{grad\iota }\phi \right) grad^{2}\iota ,
\label{28}
\end{equation}%
and defining the \emph{total specific enthalpy} $h$ of the Korteweg's fluid
as%
\begin{equation}
h=\frac{1}{2}q^{2}+\xi ,  \label{29}
\end{equation}%
we obtain (\ref{4}) and Proposition 1 is proven. The corollaries are
immediate consequences.

\section{General complex fluids.}

We follow here the unified setting proposed in [4] (and further discussed in
[15] adding the treatment of singular surfaces) with the aim to encompass a
class of complex flows as large as possible. We then consider each material
element as a subsystem collapsed into a place $\mathbf{x}$ and described by
a coarse grained order parameter $\mathbf{\nu }$. We do not specify $\mathbf{%
\nu }$; we require only that it belongs to a differentiable paracompact
manifold $\mathcal{M}$ without boundary where we presume that physical
circumstances induce a single choice of metric and of connection. We have
then a sufficiently smooth mapping $\mathcal{B}\ni \mathbf{x}\overset{%
\mathbf{\tilde{\nu}}}{\mathbf{\longmapsto }}\mathbf{\nu =\tilde{\nu}}\left( 
\mathbf{x}\right) \mathbf{\in }\mathcal{M}$.

We account for gradient effects, i.e., we consider the interaction between
neighboring material elements under the suggestions of the results about
Korteweg's fluid. These weakly non-local effects are the crucial mechanism
altering the standard thermomechanical behavior of the vorticity.

Under previous assumptions, the Lagrangian density of a general complex
fluid can be written as%
\begin{equation}
\frac{1}{2}\rho \left\vert \mathbf{v}\right\vert ^{2}+\chi \left( \mathbf{%
\nu },\mathbf{\dot{\nu}}\right) -\rho \phi \left( \iota ,\mathbf{\nu },grad%
\mathbf{\nu },\eta \right)  \label{30}
\end{equation}%
where the substructural kinetic co-energy $\chi $ (if perceptible in
experiments) depends now on $\mathbf{\nu }$ and its rate, and the potential $%
\phi $ accounts for weak non-local gradient effects. The contemporary
presence of $\mathbf{\nu }$ and $grad\mathbf{\nu }$ in the constitutive list
of entries of $\phi $\ is not simply matter of choice of generality in the
modeling (due, say, to a principle of equipresence), rather it has geometric
origin. The manifold $\mathcal{M}$ is in general not trivial (in most cases
of physical interest it is not a linear space); the pair $\left( \mathbf{%
\tilde{\nu}},grad\mathbf{\tilde{\nu}}\right) $ collects the peculiar
elements of the tangent mapping $T\mathbf{\tilde{\nu}}:T\mathcal{B}%
\rightarrow T\mathcal{M}$ between the tangent bundle of $\mathcal{B}$ and
the one of $\mathcal{M}$. In general, the two elements cannot be separated
invariantly unless $\mathcal{M}$ is endowed with a parallelism. One asks
also a physically significant parallelism. Even when $\mathcal{M}$ is
Riemannian, there is no immediate physical reason, in principle, prescribing
that the consequent Levi-Civita connection over $\mathcal{M}$ have a
prevalent r\^{o}le even in exotic cases. Taking into account the possible \
articulated geometry of $\mathcal{M}$, in general we account for both $%
\mathbf{\nu }$ and $grad\mathbf{\nu }$ as entries of $\phi $ when we decide
to model interactions between neighboring material elements.

We list just below some well known special cases of our treatment to clarify
the topic; other cases are collected at the end of this section.

\begin{enumerate}
\item For nematic liquid crystals, $\mathbf{\nu }$\ is identified with a
unit vector, say $\mathbf{n}$, indicating the direction of the prevailing
nematic order of stick molecules at $\mathbf{x}$; $\mathcal{M}$ is the unit
sphere with the identification of the antipodes and $\phi $ Oseen-Frank
potential (see e.g. [16]).

\item For semi-dilute polymeric fluids, the material element is like a box
containing a population of polymer chains described by end-to-end vectors $%
\mathbf{r}$ and $\mathbf{\nu }$ is identified at each $\mathbf{x}$ with a
second order tensor, $\mathcal{R}$, precisely the second order moment of the
distribution of the $\mathbf{r}$'s averaged over the population considered
(see e.g. [17]).

\item The simplest reasonable form of $\phi $ is of Ginzburg-Landau's type,
namely $\phi =\gamma \left( \iota ,\mathbf{\nu },\eta \right) +\frac{1}{2}%
a\left\Vert grad\mathbf{\nu }\right\Vert ^{2}$, with $a$ some appropriate
constant and $\left\Vert grad\mathbf{\nu }\right\Vert $ the norm of $grad%
\mathbf{\tilde{\nu}}\left( x\right) $ in $Hom\left( T_{\mathbf{x}}\mathcal{B}%
,T_{\mathbf{\nu }}\mathcal{M}\right) $, the space of linear maps between the
tangent space of $\mathcal{B}$ at $\mathbf{x}$ and the tangent space of $%
\mathcal{M}$ at $\mathbf{\nu =\tilde{\nu}}\left( \mathbf{x}\right) $.
\end{enumerate}

Notice that if we impose the internal constraint $\tilde{\iota}\left( 
\mathbf{x}\right) =\mathbf{\tilde{\nu}}\left( \mathbf{x}\right) $ we reduce
the treatment to the Korteweg's fluid.

\subsection{The main result about complex fluids.}

By taking into account the structure (\ref{30}) of the Lagrangian density,
we may formulate the natural generalization of Proposition 1.

\begin{proposition}
For a steady-state non-isentropic flow of a general perfect complex fluid,
the following pointwise relation holds:%
\begin{eqnarray}
\mathbf{\omega }\times \mathbf{v} &=&\vartheta grad\eta -gradh_{c}-\left(
grad\partial _{grad\mathbf{\nu }}\phi \right) ^{T}grad\mathbf{\nu }-  \notag
\\
&&-\left( grad\left( div\partial _{grad\mathbf{\nu }}\phi -\frac{d}{dt}%
\left( \partial _{\mathbf{\dot{\nu}}}\chi \right) -\partial _{\mathbf{\nu }%
}\chi \right) \right) ^{T}\mathbf{\nu }+  \notag \\
&&+\iota \left( grad\mathbf{\nu }\right) ^{T}div\partial _{grad\mathbf{\nu }%
}\phi +\iota \left( \partial _{grad\mathbf{\nu }}\phi \right) ^{T}gradgrad%
\mathbf{\nu },  \label{31}
\end{eqnarray}%
where $h_{c}$ is the total enthalpy of the complex fluid defined below.
\end{proposition}

By indicating with $\alpha $ components on some chart over $\mathcal{M}$ and
with $i,j,...$ the standard spatial components, (\ref{31}) may be read as%
\begin{eqnarray}
\left( \mathbf{\omega }\times \mathbf{v}\right) _{i} &=&\vartheta \left(
grad\eta \right) _{i}-\left( gradh_{c}\right) _{i}-  \notag \\
&&-\left( grad\left( div\partial _{grad\mathbf{\nu }}\phi -\frac{d}{dt}%
\left( \partial _{\mathbf{\dot{\nu}}}\chi \right) -\partial _{\mathbf{\nu }%
}\chi \right) \right) _{i\alpha }\mathbf{\nu }^{\alpha }-  \notag \\
&&-\left( grad\partial _{grad\mathbf{\nu }}\phi \right) _{i\alpha
}^{j}\left( grad\mathbf{\nu }\right) _{j}^{\alpha }+\iota \left( grad\mathbf{%
\nu }\right) _{i}^{\alpha }\left( div\partial _{grad\mathbf{\nu }}\phi
\right) _{\alpha j}^{j}+  \notag \\
&&+\iota \left( \partial _{grad\mathbf{\nu }}\phi \right) _{\alpha
}^{j}\left( gradgrad\mathbf{\nu }\right) _{ji}^{\alpha }.  \label{32}
\end{eqnarray}

\textbf{Corollary 3}. \emph{In a perfect complex fluid, the presence of
material substructure cancels the thermodynamic generation of vorticity when}%
\begin{eqnarray}
\vartheta grad\eta -gradh_{c} &=&\left( grad\partial _{grad\mathbf{\nu }%
}\phi \right) ^{T}grad\mathbf{\nu }-\iota \left( grad\mathbf{\nu }\right)
^{T}div\partial _{grad\mathbf{\nu }}\phi +  \notag \\
&&+\left( grad\left( div\partial _{grad\mathbf{\nu }}\phi -\frac{d}{dt}%
\left( \partial _{\mathbf{\dot{\nu}}}\chi \right) -\partial _{\mathbf{\nu }%
}\chi \right) \right) ^{T}\mathbf{\nu }-  \notag \\
&&-\iota \left( \partial _{grad\mathbf{\nu }}\phi \right) ^{T}gradgrad%
\mathbf{\nu }.  \label{33}
\end{eqnarray}

\textbf{Corollary 4}. \emph{In the case in which both the specific entropy
and the total enthalpy are constant throughout the flow of a perfect complex
fluid, the presence of material substructure may generate vorticity:}%
\begin{eqnarray}
\mathbf{\omega }\times \mathbf{v} &=&-\left( grad\partial _{grad\mathbf{\nu }%
}\phi \right) ^{T}grad\mathbf{\nu }+\iota \left( grad\mathbf{\nu }\right)
^{T}div\partial _{grad\mathbf{\nu }}\phi +  \notag \\
&&+\iota \left( \partial _{grad\mathbf{\nu }}\phi \right) ^{T}gradgrad%
\mathbf{\nu }-  \notag \\
&&-\left( grad\left( div\partial _{grad\mathbf{\nu }}\phi -\frac{d}{dt}%
\left( \partial _{\mathbf{\dot{\nu}}}\chi \right) -\partial _{\mathbf{\nu }%
}\chi \right) \right) ^{T}\mathbf{\nu }.  \label{34}
\end{eqnarray}

\subsection{Proof of Proposition 2.}

\subsubsection{Flows of isentropic perfect complex fluids.}

To obtain the field equations for general complex fluids, we consider first
the total Lagrangian%
\begin{equation}
L=\int_{\mathcal{B}}\left( \frac{1}{2}\rho \left\vert \mathbf{v}\right\vert
^{2}+\chi \left( \mathbf{\nu },\mathbf{\dot{\nu}}\right) -\rho \phi \left(
\iota ,\mathbf{\nu },grad\mathbf{\nu }\right) \right)  \label{35}
\end{equation}%
and set equal to zero the first total variation (in the sense of (\ref{10}))
of its integral over the time interval $\left[ 0,\bar{t}\right] $, namely%
\begin{equation}
\hat{\delta}\int_{0}^{\bar{t}}L=0.  \label{36}
\end{equation}%
We consider also that%
\begin{equation}
\hat{\delta}\mathbf{\nu =}\delta \mathbf{\nu }+\left( grad\mathbf{\nu }%
\right) \delta \mathbf{x},  \label{37}
\end{equation}%
in components%
\begin{equation}
\left( \hat{\delta}\mathbf{\nu }\right) ^{\alpha }\mathbf{=}\left( \delta 
\mathbf{\nu }\right) ^{\alpha }+\left( grad\mathbf{\nu }\right) _{i}^{\alpha
}\left( \delta \mathbf{x}\right) ^{i},  \label{38}
\end{equation}%
and that%
\begin{equation}
\int_{\mathcal{B}}\rho \hat{\delta}\phi =\int_{\mathcal{B}}\rho \left(
\partial _{\iota }\phi \hat{\delta}\iota +\partial _{\mathbf{\nu }}\phi
\cdot \hat{\delta}\mathbf{\nu }+\partial _{grad\mathbf{\nu }}\phi \cdot \hat{%
\delta}grad\mathbf{\nu }\right) .  \label{39}
\end{equation}

By using repeatedly Gauss theorem, variations vanishing at the boundary of $%
\mathcal{B}$ and the guidelines of Section 2.2.1 (see also [4], p. 30-34),
we obtain the following balances:%
\begin{equation}
\rho \frac{d\mathbf{v}}{dt}=div\mathbf{T}\text{ \ \ , \ \ }div\mathcal{S}-%
\mathbf{z}=\frac{d}{dt}\partial _{\mathbf{\dot{\nu}}}\chi -\partial _{%
\mathbf{\nu }}\chi ,  \label{40}
\end{equation}%
where the stress tensor $\mathbf{T}$ has in this case the form%
\begin{equation}
\mathbf{T=}\rho \iota \partial _{\iota }\phi \mathbf{I}-\left( grad\mathbf{%
\nu }\right) ^{T}\partial _{grad\mathbf{\nu }}\phi  \label{41}
\end{equation}%
and we use $\mathcal{S}$ and $\mathbf{z}$ as%
\begin{equation}
\mathcal{S}=\rho \partial _{grad\mathbf{\nu }}\phi \text{ \ \ , \ \ }\mathbf{%
z=}\rho \partial _{\mathbf{\nu }}\phi ,  \label{42}
\end{equation}%
calling them microstress and self-interaction, respectively. The pointwise
balance of substructural interactions (\ref{40}b) states that
self-interactions $\mathbf{z}$ accrue within each material element (due to
its substructure) to balance the `contact' interactions $\mathcal{S}$
between neighboring material elements (due to the relative change of
substructure). At each $\mathbf{x}$, $\mathbf{z}$ is an element of the
cotangent space $T_{\mathbf{\nu }}^{\ast }\mathcal{M}$ while $\mathcal{S}$
belongs to $Hom\left( T_{\mathbf{x}}^{\ast }\mathcal{B},T_{\mathbf{\nu }%
}^{\ast }\mathcal{M}\right) $. As for the pair $\left( \mathbf{\nu },grad%
\mathbf{\nu }\right) $, even in this case the non-trivial structure of $%
\mathcal{M}$ does not allow us to separate invariantly the microstress from
the self-interaction, in general. However, when an appropriate parallelism
permits us to select an explicit form of $\phi $\ depending on the sole $grad%
\mathbf{\nu }$\ and substructural inertia effects are negligible, equation (%
\ref{40}b) reduces to a divergence form. For example, such a situation
happens in the special case of nematic liquid crystals when we select just
the simplest Frank's potential and reduce ourselves to investigate only the
physics described by harmonic maps taking values in $S^{2}$ [19].

As anticipated in discussing the special case of Korteweg's fluid, the
kinetic co-energy $\chi \left( \mathbf{\nu },\mathbf{\dot{\nu}}\right) $\ is
such that its Legendre transform with respect to the rate $\mathbf{\dot{\nu}}
$\ coincides with the substructural kinetic energy $k\left( \mathbf{\nu },%
\mathbf{\dot{\nu}}\right) $, i.e., $k=\partial _{\mathbf{\dot{\nu}}}\chi
\cdot \mathbf{\dot{\nu}}-\chi $. Its contribution may be in general
sensible. Notice that, even when $k\left( \mathbf{\nu },\mathbf{\dot{\nu}}%
\right) $ is of the form $\frac{1}{2}\dot{\nu}^{\alpha }\Omega _{\alpha
\beta }\dot{\nu}^{\beta }$, the explicit form of $\chi $ differs from $k$ by
an addendum of the type $\lambda _{\alpha }\dot{\nu}^{\alpha }$, which may
be helpful to describe some prominent physical effects. For example, in the
case of magnetizable fluids, if we select the magnetization vector as order
parameter and constraint its rate to be of gyroscopic nature, we reduce (\ref%
{40}b) to the equation of Gilbert.

\subsubsection{Details about Proposition 2.}

When the flow is not isentropic, the potential is of the form%
\begin{equation}
\phi =\phi \left( \iota ,\mathbf{\nu },grad\mathbf{\nu },\eta \right)
\label{43}
\end{equation}%
and we define (as in (\ref{24})) the specific enthalpy density $\xi _{c}$ of
a complex fluid as the opposite of the partial Legendre transform with
respect to the morphological descriptors, namely%
\begin{equation}
\xi _{c}=\phi -\iota \partial _{\iota }\phi -\partial _{\mathbf{\nu }}\phi
\cdot \mathbf{\nu }-\partial _{grad\mathbf{\nu }}\phi \cdot grad\mathbf{\nu .%
}  \label{44}
\end{equation}%
We may also rewrite the term \thinspace $-\iota \partial _{\iota }\phi $ as $%
\iota \tilde{p}$, where $\tilde{p}$ is the pressure relevant for this case,
given by $-\partial _{\iota }\phi $. Equation (\ref{22}) becomes%
\begin{equation}
\mathbf{\omega }\times \mathbf{v}=-\frac{1}{2}gradq^{2}-\iota grad\tilde{p}%
-\iota div\mathbf{\bar{T}}^{E},  \label{45}
\end{equation}%
with%
\begin{equation}
\mathbf{\bar{T}}^{E}=\left( grad\mathbf{\nu }\right) ^{T}\partial _{grad%
\mathbf{\nu }}\phi \text{ \ \ , \ \ }\left( \mathbf{\bar{T}}^{E}\right)
_{i}^{j}=\left( grad\mathbf{\nu }\right) _{i}^{\alpha }\left( \partial _{grad%
\mathbf{\nu }}\phi \right) _{\alpha }^{j}.  \label{46}
\end{equation}%
Then we get%
\begin{equation}
div\mathbf{\bar{T}}^{E}=\left( grad\mathbf{\nu }\right) ^{T}div\partial
_{grad\mathbf{\nu }}\phi +\left( \partial _{grad\mathbf{\nu }}\phi \right)
^{T}gradgrad\mathbf{\nu }.  \label{47}
\end{equation}

The gradient of (\ref{44}) allows us to write also%
\begin{equation}
-\iota grad\tilde{p}=\vartheta grad\eta -grad\xi _{c}-\left( grad\partial _{%
\mathbf{\nu }}\phi \right) ^{T}\mathbf{\nu }-\left( grad\partial _{grad%
\mathbf{\nu }}\phi \right) ^{T}grad\mathbf{\nu }\text{.}  \label{48}
\end{equation}%
By inserting (\ref{48}) and (\ref{47}) into (\ref{45}) and taking into
account (\ref{40}b) and (\ref{42}b), we get Proposition 2 where%
\begin{equation}
h_{c}=\frac{1}{2}q^{2}+\xi _{c}.  \label{49}
\end{equation}

The corollaries are immediate.

\subsection{Further special cases}

Well-known cases covered by our general modeling have been listed at the
beginning of Section 3, namely, nematic liquid crystals and polymeric
fluids, and we may also mention ferrofluids once, in presence of external
electric fields, we add to $\phi $ the electromagnetic energy and develop
appropriate variations of the relevant fields.

We may also investigate more exotic cases by making use of the flexibility
of the general unifying framework in which we act. Some examples are listed
below, but they do not exhaust the range of possibilities.

\begin{enumerate}
\item (Smectic-A liquid crystals.) In the smectic-A phase, liquid crystals
appear organized in layers in which stick molecules tend to be aligned
orthogonally to the layer interface. Within each layer the molecules flow
freely; in the orthogonal direction (where the behavior is basically the one
of a one-dimensional crystal) they may permeate from a layer into another.
Linear [20] and non-linear [21-23] continuum models have been formulated to
describe various aspects of the elasticity and hydrodynamics of smectic-A
phase, paying a few attention on the behavior of possible vortices there.
Since the current formulation of the hydrodynamics of smectic-A liquid
crystals falls within the general setting discussed here, under the
conditions of Proposition 2 and subsequent corollaries, we could foresee
formation or cancellation of vortices. In fact, natural ingredients for
modeling the smectic-A phase are a vector $\mathbf{n}\in S^{2}$ at each $%
\mathbf{x}\in \mathcal{B}$ describing the local orientational order, and a
scalar function $w$ parametrizing the layers. More precisely, we may write
[23] $w\left( \mathbf{x},t,a\lambda \right) $, with $\lambda $\ an
appropriate length scale and $a$ running in a set of integers, to account
for possible unequal spacing of layers. However, at a (gross) coarse grained
scale, we may imagine that $w\left( \mathbf{x},t,\cdot \right) $ be defined
in a continuum approximation on an interval of the real line so that, as
shown in [23], $\left\vert gradw\right\vert ^{-1}$ measures the current
thickness of the layers. Then, we can select the order parameter as the pair 
$\left( \mathbf{n},w\right) $; however, in absence of tilt (a condition
reasonable outside the defect core) we have strictly%
\begin{equation}
\mathbf{n}=\frac{gradw}{\left\vert gradw\right\vert },  \label{50}
\end{equation}%
and the simplest choice for the potential $\phi $\ in the compressible case
is of the form (see [22])%
\begin{equation}
\phi \left( \iota ,w,gradw\right) =\bar{\phi}\left( \iota \right) +\frac{1}{2%
}\gamma _{1}\left( \left\vert gradw\right\vert -1\right) ^{2}+\frac{1}{2}%
\gamma _{2}\left( div\mathbf{n}\right) ^{2},  \label{52}
\end{equation}%
with the identification (\ref{50}), and with $\gamma _{1}$\ and $\gamma _{2}$%
\ material constants. In (\ref{52}), the term $\left( \left\vert
gradw\right\vert -1\right) ^{2}$ accounts for the compression of layers
while $\left( div\mathbf{n}\right) ^{2}$ describes the nematic phase and is
the first addendum of (three constant) Frank's potential (see [3], p. 55).
Inertial effects can be considered by putting $\chi =\frac{1}{2}\rho \alpha
\left\vert \mathbf{\dot{n}}\right\vert ^{2}$, with $\alpha $ a material
constant [23]. The special counterparts of (\ref{31}), (\ref{33}) and (\ref%
{34}), associated with (\ref{52}), follow after straightforward
calculations. To simplify further the matter, one may imagine to have an
incompressible flow, neglecting $\bar{\phi}\left( \iota \right) $, a
circumstance not incompatible with the assumed compressibility of the
layers. Taking into account the incompressible limit of (\ref{52}), namely $%
\phi \left( w,gradw\right) $, the application of Proposition 2, in absence
of prominent substructural inertia of stick molecules in smectic-A phase,
leads to%
\begin{eqnarray}
\mathbf{\omega }\times \mathbf{v} &=&\vartheta grad\eta -gradh_{c}-\left(
grad\mathcal{S}\right) ^{T}gradw-  \notag \\
&&-wgrad\left( div\mathcal{S}\right) +\iota div\mathcal{S}\left(
gradw\right) +\iota \mathcal{S}\left( gradgradw\right) ,  \label{52bis}
\end{eqnarray}%
with%
\begin{equation}
\mathcal{S}=\gamma _{1}\left( \left\vert gradw\right\vert -1\right) \mathbf{n%
}-\gamma _{2}\left\vert gradw\right\vert ^{-1}\left( \mathbf{I}-\mathbf{%
n\otimes n}\right) grad\left( div\mathbf{n}\right) ,  \label{52ter}
\end{equation}%
and $\mathbf{I}$ the second-order unit tensor.

\item (Polyelectrolyte polymers and polymer stars.) In the case of
polyelectrolyte polymers suffering possible polarization, we may imagine to
assign at each point $\mathbf{x}\in \mathcal{B}$ not only the dyadic tensor $%
\mathcal{R}$, mentioned at the beginning of Section 3, but also a
polarization vector $\mathbf{p}$. By indicating with $Vec$ the translation
space over $\mathcal{E}$, and with $B_{p}$ a ball of $Vec$ centered at zero
and with radius equal to $p$, the maximum amplitude of the polarization, we
have $\mathcal{M}=Hom\left( Vec,Vec\right) \times B_{p}$ and we may
construct the relevant hydrodynamic theory, once an explicit expression for $%
\phi $ has been selected (see examples in [17]), adding, in presence of
external electric fields, the electromagnetic energy. When we consider also
polymer stars, the picture becomes more articulated and we may imagine to
have $\mathcal{M}=Hom\left( Vec,Vec\right) \times B_{p}\times \left[ 0,b%
\right] $. In this case, at each $\mathbf{x}\in \mathcal{B}$, an arbitrary
element of the interval $\left[ 0,b\right] $ of the real line describes the 
\emph{radius of gyration} of the polymeric molecules.

\item (Superfluid helium $^{4}He$.) The work of Rasetti and Regge [24]
allows us to manage suitable Poisson brackets to describe `directly' the
dynamics of vortices in superfluid helium (see also the developments in
[25]). However, following the suggestions of [4] (p. 7), we may choose to
adopt the general point of view used here and to select $\mathcal{M}%
=S^{1}\subset \mathbb{C}$, using in this way a potential $\phi $ with a
Ginzburg-Landau structure with complex entries. The application of
Proposition 2 and subsequent corollaries follows.
\end{enumerate}

\section{Concluding remarks.}

In the case of Korteweg's fluid, diffuse interfaces modeled by the specific
volume may generate an energetic wall against the thermodynamic formation of
vorticity. However, there are conditions for which such an excess of energy
may favour vorticity even for an isentropic flow.

Analogous phenomena occur in the case of general complex fluids. Notice
that, excluding possible inertial terms associated with $\chi $, the
gradient of $\mathbf{\nu }$\ and the derivative of $\phi $\ with respect to $%
grad\mathbf{\nu }$ are prominent ingredients into the vorticity equation (%
\ref{31}). This circumstance suggests that inhomogeneous spatial
distributions of substructures (preferably with high gradients) in
steady-state flows may generate energetic barriers to vorticity or may
favour it. Basically, the energy involved is the one associated with weakly
non-local interactions between neighboring material elements.

For non-stationary incompressible two-dimensional flows the vorticity is in
general not conserved even in the perfect case due to \emph{a mechanism of
transfer of energy between the coarse grained level and the substructural one%
}, ruled by the tensor $\mathbf{T}^{E}$. In fact, if we calculate the $curl$
of (\ref{40}a) taking into account (\ref{41}), we get%
\begin{equation}
\frac{d\mathbf{\omega }}{dt}=\left( \mathbf{\omega }\cdot grad\right) 
\mathbf{v}+grad\iota \times div\mathbf{T}-\iota curl\left( div\mathbf{T}%
^{E}\right) .  \label{55}
\end{equation}%
When the fluid is incompressible, the addendum $grad\iota \times div\mathbf{T%
}$ vanishes, and in a two-dimensional flow we have also $\left( \mathbf{%
\omega }\cdot grad\right) \mathbf{v}=0$, but the vorticity is still altered
by the term $-\iota curl\left( div\mathbf{T}^{E}\right) $, which is
different from zero except when (i) there exists a scalar field $\mathcal{B}%
\ni \mathbf{x}\longmapsto \pi \left( \mathbf{x}\right) $ such that $grad\pi
=div\mathbf{T}^{E}$, or (ii) there exists a second order tensor valued field 
$\mathcal{B}\ni \mathbf{x}\longmapsto \mathbf{A}\left( \mathbf{x}\right) \in
Hom\left( Vec,Vec\right) $ such that $curl\left( \mathbf{T}^{E}\right)
^{T}=\left( curl\mathbf{A}\right) ^{T}$. So that, in time-dependent states,
even in a two-dimensional flow of an \emph{incompressible} perfect complex
fluid the vorticity can be altered with consequent generation of topological
transitions.

Experiments confirming the theoretical predictions of the present paper in
some special case could be useful toward the understanding of complex fluids.

\ \ \ \ \ \ 

\textbf{Acknowledgements}. I thank two anonymous Referees for stimulating
suggestions. The support of the Italian National Group of Mathematical
Physics (GNFM-INDAM) is acknowledged.

\section{References}

\begin{enumerate}
\item Crocco, L. (1937), Eine neue Stromfunktion f\"{u}r die Erforschung der
Bewegung der Gase mit Rotation, \emph{Z. angew. Math. Mech.}, \textbf{17},
1-7.

\item Vaszonyi, A. (1945), On rotational gas flows, \emph{Q. Appl. Math.}, 
\textbf{3}, 29-37.

\item Serrin, J. (1959), Mathematical principles of classical fluid
mechanics, \emph{Handbuch der Physik}, \textbf{8}$_{1}$, Springer Verlag,
Berlin, 125-363.

\item Capriz, G., \emph{Continua with microstructure}, Springer Verlag,
Berlin, 1989.

\item Korteweg, D. J. (1901), Sur la forme que prennent les \'{e}quations du
movement des fluides si l'on tient compte des forces capillaires caus\'{e}es
par des variations de densit\'{e} consid\'{e}rables mais continues et sur la
th\'{e}orie de la capillarit\'{e} dans l'hipoth\`{e}se d'une variation
continue de la densit\'{e}, \emph{Arch. N\'{e}erl. Sci. Exactes Nat. Ser. }%
II,\emph{\ }\textbf{6}, 1-24.

\item Joseph, D. D. (1990), Fluid dynamics of two miscible liquids with
diffusion and gradient stresses, \emph{Eur. J Mech. B / Fluids}, \textbf{9},
565-596.

\item Chen, C.-Y. and Meiburg, E. (2002), Miscible displacements in
capillary tubes: influence of Korteweg stresses and divergence effects, 
\emph{Phys. Fluids}, \textbf{14}, 2052-2058.

\item Chen, C.-Y., Wang, L. and Meiburg, E. (2001), Miscible droplets in a
porous medium and the effects of Korteweg stresses, \emph{Phys. Fluids}, 
\textbf{13}, 2447-2456.

\item Galdi, G. P., Joseph, D. D., Preziosi, L. and Rionero, S. (1991),
Mathematical problems for miscible, incompressible fluids with Korteweg
stresses, \emph{Eur. J Mech. B / Fluids}, \textbf{10}, 253-267.

\item Danchin, R. and Dejardins, B. (2001), Existence of solutions for
compressible fluid models of Korteweg type, \emph{Ann. Inst. Henri Poincar%
\'{e} Anal. nonlineaire}, \textbf{18}, 97-133.

\item Rowlinson, J. S. and Widom, B., \emph{Molecular theory of capillarity}%
, Clarendon Press, Oxford, 1989.

\item Dunn, J. E. and Serrin, J. (1985), On the thermodynamics of
interstitial working, \emph{Arch. Rational Mech. Anal.}, \textbf{88}, 95-133.

\item Capriz, G. (1995), Continua with latent microstructure, \emph{Arch.
Rational Mech. Anal.}, \textbf{90}, 43-56.

\item Anderson, D. M., McFadden, G. B. and Wheeler, A. A. (1998), Diffuse
interface methods in fluid mechanics, \emph{Ann. Rev. Fluid Mech.}, \textbf{%
30}, 139-165.

\item Mariano, P. M. (2001), Multifield theories in mechanics of solids, 
\emph{Adv. Appl. Mech}., \textbf{38}, 1-93.

\item De Gennes, P.-G., \emph{Introduction to polymer dynamics}, Cambridge
University Press, Cambridge, 1990.

\item Likos, C. N. (2001), Effective interactions in soft condensed matter
physics, \emph{Physics Reports}, \textbf{348}, 267-439.

\item DiBattista, M. T., Majda, A. J. and Grote, M. J. (2001),
Meta-stability of equilibrium statistical structures for prototype
geophisical flows with damping and driving, \emph{Physica D}, \textbf{151},
271-304.

\item Brezis, H., Coron, J.-M. and Lieb, E. H. (1986), Harmonic maps with
defects, \emph{Comm. Math. Phys.}, \textbf{107}, 649-705.

\item De Gennes, P.-G. and Prost, J., \emph{The physics of liquid crystals},
Oxford University Press, Oxford, 1993.

\item Capriz, G. (1995), Smectic liquid crystals as continua with latent
microstructure, \emph{Meccanica}, \textbf{30}, 621-627.

\item E, W. (1997), Nonlinear continuum theory of smectic-A liquid crystals, 
\emph{Arch. Rational Mech. Anal.}, \textbf{137}, 159-175.

\item Capriz, G. and Napoli, G. (2001), Swelling and tilting in smectic
layers, \emph{Appl. Math. Lett.}, \textbf{14}, 673-678.

\item Rasetti, M. and Regge, T. (1975), Vortices in He-II, current algebras
and quantum knots, \emph{Physica A}, \textbf{80}, 217-223.

\item Holm, D. D. (2003), Rasetti-Regge Dirac bracket formulation of
Lagrangian fluid dynamics of vortex filaments, \emph{Math. Comp. Simulation}%
. \textbf{62}, 53-63.
\end{enumerate}

\end{document}